\documentclass[preprint,english,12pt]{article}
\usepackage{amsmath}
\usepackage{esint}
\usepackage{amssymb}
\usepackage{graphicx, epsfig}
\usepackage{mathrsfs}
\usepackage{hyperref}
\usepackage{bm}
\usepackage {amsmath}
\makeatletter
\usepackage{babel}
\makeatother
\allowdisplaybreaks

\begin{document}

\title{\bf{A dark energy model alternative to generalized Chaplygin gas}}

\author{Hoavo Hova\footnote{hovhoav@mail.ustc.edu.cn} 
and Huan-Xiong Yang\footnote{hyang@ustc.edu.cn}\\
Interdisciplinary Center for Theoretical Study,\\ 
University of Science and Technology of China, \\
Hefei, 230026, China\\
{}\\
}

\date{}

\maketitle

\vspace{-8.cm}
\begin{center}
\hfill USTC-ICTS-10-20
\end{center}
\vspace{8.cm}

\begin{abstract} 
We propose a new fluid model of dark energy for $-1 \leq \omega_{\text{eff}} \leq 0$ 
    as an alternative to the generalized Chaplygin gas models. The energy density of dark 
    energy fluid is severely suppressed during barotropic matter dominant epochs, and it
    dominates the universe evolution only for eras of small redshift. From the perspective 
    of fundamental physics, the fluid is a tachyon field with a scalar potential flatter 
    than that of power-law decelerated expansion. Different from the standard 
    $\Lambda\text{CDM}$ model, the suggested dark energy model claims that the
    cosmic acceleration at present epoch can not continue forever but will cease in 
    the near future and a decelerated cosmic expansion will recover afterwards. 
\end{abstract}

\section{Introduction}

Recent cosmic observations, including Type Ia Supernovae\cite{Riess, 
Perlmutter}, Large Scale Structure (LSS)\cite{Tegmark1, Tegmark2, 
Tegmark3, Seljak, Adelman, Abazajian} and Cosmic Microwave Background
(CMB)\cite{Spergel1, Page, Hinshaw, Jarosik} have independently 
indicated that the universe is currently undergoing an accelerated 
expansion phase. The source for this late time cosmic acceleration
has been dubbed ``dark energy'', which is distinguished from ordinary 
matter species such as baryons and radiation in the sense that it has 
a negative pressure. The cosmic accelerated expansion of the Universe
occurs when the gravitational force produced by the barotropic ordinary 
matter is counteracted by the negative pressure of dark energy. Despite
many years of research and much progress, the nature and the origin of 
dark energy have not been confirmed yet.

Phenomenologically, the best candidate of dark energy that is perfectly 
consistent with the current observation data is the so-called cosmological 
constant (CC) $\Lambda$ which was introduced by Einstein in his 
gravitational field equations\cite{Kunz}. However, CC explanation for 
dark energy encounters some fundamental obstacles in physics, \emph{e.g.}, 
the fine-tuning problem and the coincidence problem\cite{Amendola1}. From 
the particle physics perspective, the CC should be interpreted as the 
energy density $\rho_{vac}$ of the quantum vacuum, which is close to 
Planck density $M_{P}^{4}$ ($M_{P}=1/\sqrt{8\pi G}$ is the reduced Planck 
mass) in magnitude. The observed value of the dark energy density is much 
less than that, $\rho_{obs} \approx 10^{-123}\rho_{vac}$. Eliminating this
great difference of about 123 orders of magnitude between the observed
value of dark energy and that estimated from quantum field theory requires 
some severe fine-tuning mechanisms to work\cite{Hawking, Kachru, Tye, 
Yokoyama, Mukohyama, Kane, Dolgov}. On the other hand, even if this 
fine-tuning problem could be evaded, the coincidence problem remains due 
to the fact that the vacuum energy is time independent and non-dynamical.

The resolution of CC problems has probably to wait for the advent of a 
satisfied quantum theory of gravity in 4-dimensional spacetime. As a 
temporary expedient in cosmology community, the fine-tuning problem is
assumed to have been solved for some underlying reasons: CC is zero for
some reason and dark energy that drives the late time cosmic accelerated 
expansion has nothing to do with it. Notice that the observations actually 
say little about the time evolution of the equation of state (EoS) of 
dark energy, researchers have proposed many alternative models (see Review 
article \cite{Copeland} and references listed therein) to explain the late 
time cosmic acceleration and alleviate the corresponding coincidence
problem, among which the Chaplygin gas model \cite{Kamenshchik} appears 
very interesting. In such a simple model, the acceleration of the universe
at late times is due to an alien fluid that has the specific EoS $p=-A/\rho$,
with $A$ a positive constant. The intriguing characteristic of Chaplygin 
gas is that in the flat Robertson-Walker background, $ds^{2} = - dt^{2}
+ a^{2}(t)d\bm{x}^{2}$ ($a$ is the scale factor of an expanding universe),
\begin{equation}\label{eq:1}
\rho = \sqrt{A + \frac{B}{a^6}}
\end{equation}
where $B$ is an integration constant. At early times when $a \ll (B/A)^{1/6}$,
the gas behaves as a pressureless dust, $\rho \sim \sqrt{B}/a^3$. Meanwhile 
it approaches asymptotically a cosmological constant ($\rho \sim -p \sim
\sqrt{A}$) at late times when $a \gg (B/A)^{1/6}$. The Chaplygin gas model
of dark energy suffers from strong observational pressure from explaining CMB
anisotropies \cite{Amendola2}. This shortcoming is alleviated in a generalized
Chaplygin gas model introduced in \cite{Bento} with $p = - A/\rho^{\alpha}$. 
However, it follows from the observational data that the parameter $\alpha$
is severely constrained within the range $0 \leq \alpha \leq 0.2$ at the 
$95\%$ confidence level, implying the little difference between the time 
dependent generalized Chaplygin gas and the  time independent cosmological 
constant.

In this paper we study a new kind of fluid models of dark energy, whose EoS 
is phenomenologically suggested to be
\begin{equation}
  \label{eq:2}
  p = - \rho + \rho \text{sinc}( \mu \pi \rho^{(0)}/\rho)
\end{equation}
where $\text{sinc}(\xi):=\sin(\xi)/\xi$, $\mu$ is a dimensionless fine-tuning
parameter, and $\rho^{(0)}$ is assumed to be the present energy density of 
the dark energy fluid. Our aim is to examine to what extent this fluid model 
can deviate from the time independent CC. The investigation is restricted at 
the background level. We find that during the matter era the energy density
of suggested dark energy fluid is suppressed compared to that of dark matter, 
satisfying the requirement for the sufficient growth of large-scale structure 
and the constraint of the universe age problem. The onset of late time cosmic
acceleration occurs around the redshift $z\sim 0.6$. The coincidence problem
why the accelerated expansion of the universe is rephrased as why the EoS of
dark energy fluid takes the suggested form, for which we have proposed a tachyon
interpretation originated from superstring theory. Throughout the paper we 
adopt the Planck units $c= \hbar = M_{P}=1$, and assume that there is no direct 
interactions between different fluids.

\section{Model}

EoS of the pure CC is $p= - \rho$, where both $p$ and $\rho$ are constant as
we can see from the energy conservation equation when dark energy does not 
couple with barotropic fluids. In obtaining a dynamical dark energy fluid
that deviates from pure CC, we can simply amend the EoS of such a fluid to 
$p = -\rho + F(\rho)$ with an energy density dependent function $F(\rho)$.
Since lots of astrophysical data support the pure CC behaviour of dark energy 
fluid in the present time, we choose
\begin{equation}
  \label{eq:3}
  F(\rho) = \rho \text{sinc}( \mu \pi \rho^{(0)}/\rho)
\end{equation}
EoS of the supposed dark energy fluid can alternatively be described by the
ratio $\omega = p/\rho$, which turns out to be
\begin{equation}
  \label{eq:4}
  \omega = -1 + \text{sinc} \big( {\mu \pi \rho^{(0)}}/{\rho} \big)
\end{equation}
In view of the astrophysical constraint on the universe age\cite{Spergel2}, we 
simply take the value of fine-tuning parameter $\mu$ to be $\mu \approx 0.931$ 
in this work (See explanation below Eq.(\ref{eq:21})). Thus, at large redshift, 
$\rho \gg \rho^{(0)}$, $\omega \rightarrow 0$. While at present, $\rho \approx
\rho^{(0)}$, $\omega \approx -1$.

To obtain a viable dark energy model we have to require that the energy density
of dark energy fluid remains negligible during the barotropic matter dominating
eras, emerging only at late times to give rise to the current observed 
accelerated expansion of the universe. Therefore, we assume the coexistence of
some independent barotropic fluids with suggested dark energy fluid in the model. 
EoS of these barotropic fluids is assumed to be $\omega_{b} = p_{b}/\rho_{b}$, 
where for simplicity $\omega_{b} = \text{Const}$\footnote{If the barotropic fluid consists only of the cold dark matter, we have 
    $\omega_{b}=\omega_{m}=0$.}. The Friedmann 
equations in the flat Robertson-Walker background read
\begin{eqnarray}
  \label{eq:5a}
  &  & 3H^{2} = \rho + \sum_{b} \rho_{b} \\
  &  & 2\dot{H} = -\rho -p - \sum_{b}(1+ \omega_{b}) \rho_{b} 
  \label{eq:5b}
\end{eqnarray}
where $H := \dot{a}/a$ is the Hubble parameter, and a dot denotes the derivative
with respect to the cosmic time $t$. The dark energy fluid is supposed not to 
interact with the barotropic fluid. Therefore, 
\begin{equation}
  \label{eq:6}
  \dot{\rho} + 3H (\rho +p) =0, ~~~~~\dot{\rho}_{b} 
  + 3H (1 + \omega_{b}) \rho_{b} = 0.
\end{equation}
The solutions to these two equations are 
\begin{equation}
  \label{eq:7}
 \rho = \frac{\mu \pi \rho^{(0)}}{2\arctan[a^{3}\tan({\mu \pi}/2)]}
\end{equation}
and
\begin{equation}
  \label{eq:8}
  \rho_{b} = \frac{\rho_{b}^{(0)}} {a^{3(1+ \omega_{b})}}
\end{equation}
respectively. Notice that for radiation $\omega_b=\omega_\gamma=1/3$ but for a 
non-relativistic matter $\omega_{b}=\omega_m=0$. To examine whether the onset 
of the late time cosmic acceleration occurs around the redshift $z\sim 1$, we 
have to study the variation of dimensionless energy density of suggested dark 
energy fluid
\begin{equation}
  \label{eq:9}
  \Omega := \frac{\rho}{3H^{2}}
\end{equation}
with respect to the redshift variable $z$, which is related to the scale 
factor $a(t)$ by relation $a = 1/(1+z)$. Using $z$, the Friedmann equation
(\ref{eq:5a}) becomes
\[
3H^{2}(z) = \rho + \sum_{b} \rho_{b}^{(0)} (1+z)^{3(1+ \omega_{b})}
\]
which implies,
\begin{equation}
  \label{eq:10}
  \frac{H^{2}(z)}{H^{2}_{0}} = \frac{1}{1-\Omega}
  \sum_{b}\Omega_{b}^{(0)}(1+z)^{3(1+ \omega_{b})}
\end{equation}
where $H_{0}$ is the present value of the Hubble rate,  $H_{0}^{-1} \approx 
2998 h^{-1} \text{Mpc}$ and $h \approx 0.72$. In Eq.(\ref{eq:10}) and 
afterwards we use
\begin{equation}
  \label{eq:11}
  \Omega^{(0)} = \frac{\rho^{(0)}}{3H_{0}^{2}}, 
  ~~~~ \Omega_{b}^{(0)} = \frac{\rho_{b}^{(0)}}{3H_{0}^{2}}
\end{equation}
to stand for the dimensionless energy densities of dark energy and barotropic  
fluids at present epoch respectively. Sometimes $E(z):=H(z)/H_{0}$ is called 
the  {\it{scaled Hubble expansion rate}}. With $\Omega$ and $\Omega^{(0)}$, 
Eq.(\ref{eq:7}) becomes,
\[
\frac{H^{2}}{H^{2}_{0}} = \frac{\mu \pi \Omega^{(0)}}{2 \Omega \arctan [
  (1+z)^{-3} \tan({\mu\pi}/2)]}
\]
The equivalence of this equation and Eq.(\ref{eq:10}) indicates
\begin{equation}
  \label{eq:12}
  \frac{H^{2}}{H_{0}^{2}}= \frac{\mu \pi \Omega^{(0)}}{2 \arctan [
    (1+z)^{-3}\tan ({\mu \pi}/2)] } 
    + \sum_{b}\Omega_{b}^{(0)} (1+z)^{3(1+\omega_{b})}
\end{equation}
By using the redshift variable $z$ and Eq.(\ref{eq:7}), we can rewrite the
acceleration equation (\ref{eq:5b}) as
\begin{eqnarray}
&  & \frac{d}{dz}\Big(\frac{H^{2}}{H_{0}^{2}} \Big) = \frac{3}{(1+z)}
\bigg[\frac{ \mu \pi \Omega^{(0)} (1+z)^{3}\tan(\mu\pi/2)) }{2\big((1+z)^{6} 
+ \tan^{2}(\mu\pi/2) \big) \arctan^{2}((1+z)^{-3}\tan(\mu\pi/2)) }
\nonumber \\
&  & ~~~~~~~~~~~~~~~ + \sum_{b} (1 + \omega_{b}) \Omega_{b}^{(0)}
(1+z)^{3(1+\omega_{b})}
\bigg ] 
\label{eq:13}
\end{eqnarray}
Obviously, (\ref{eq:12}) is the solution of Eq.(\ref{eq:13}) under the initial
condition $H(z)|_{z=0}=H_{0}$. The change of dimensionless energy density of 
dark energy fluid with respect to redshift $z$ is given by
\begin{equation}
  \label{eq:14}
  \Omega = \frac{\mu \pi \Omega^{(0)}}{\mu \pi \Omega^{(0)} 
  + 2 \arctan [
    (1+z)^{-3}\tan ({\mu \pi}/2)] \sum_{b}\Omega_{b}^{(0)} 
    (1+z)^{3(1+\omega_{b})}}
\end{equation}
In FIG.1, we show the $\Omega \sim z$ plot for the suggested model, where
we have assumed that the barotropic fluids consist of radiation and
non-relativistic cold dark matter, and chosen $\Omega^{(0)} \approx 0.73$,
$\Omega_{m}^{(0)} \approx 0.27$ and $\Omega_{r}^{(0)} \approx 8.1 \times
10^{-5}$ in consistent with the current observations\cite{Spergel2}. Different 
from CC, the energy density of dark energy fluid in the model under 
consideration evolves, which remains non-zero for a considerable duration so
that the coincidence problem is alleviated. The redshift dependent $\Omega$ reaches its  
maximum $\Omega_{\text{Max}}=1$ at $z=-1$ while for both $z>-1$ and 
$z<-1$ cases it decays monotonously. The suggested dark energy dominates the 
universe energy only for small redshift stages ($-2.6 \lesssim z \lesssim 0.6$). 
When $z\approx 0.59$ (or $z \approx -2.59$), $\Omega$ is caught up with by
that of cold dark matter. In the duration for $z \gg 0.59$ (or $z \ll -2.59$) 
the energy density of dark energy fluid is considerably suppressed so 
that the sufficient growth of large-scale structures in the universe is allowed. 
\begin{figure}[ht]
\begin{center}
\includegraphics{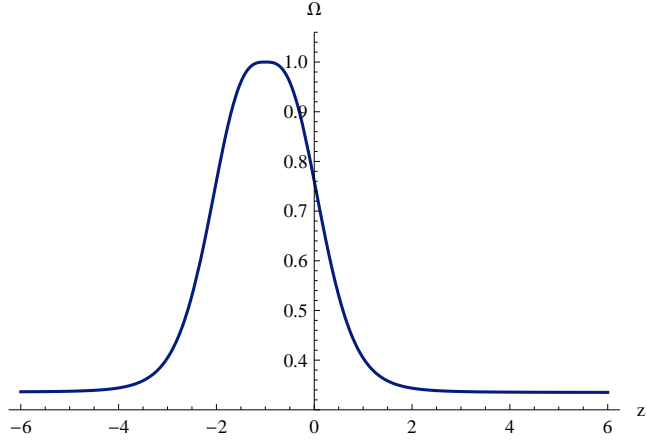}\hfill
\caption{\small{The change of $\Omega$ with respect to $z$ in the model under
consideration. We assume that the barotropic fluids consist of radiation and
non-relativistic cold dark matter, and choose $\Omega^{(0)} \approx 0.73$,
$\Omega_{m}^{(0)} \approx 0.27$ and $\Omega_{r}^{(0)} \approx 8.1 \times
10^{-5}$. When $z \approx 10$,  $\Omega \approx 0.33$ while when $z \approx 
10^{6}, \Omega \approx 0.0015$. The energy density of the suggested dark 
energy fluid decays monotonously for both $z>-1$ and $z<-1$ cases, which 
dominates only for the small redshift stage ($-2.6 \lesssim z \lesssim 0.6$).}}
\label{hy1}
\end{center}
\end{figure}

In terms of redshift $z$, EoS of the dark energy fluid defined in
Eq.(\ref{eq:4}) becomes,
\begin{equation}
  \label{eq:15}
  \omega = -1 + \frac{(1+z)^{3} \tan({\mu\pi}/2)}{[(1+z)^{6} +
    \tan^{2}({\mu\pi}/2)]\arctan\big[(1+z)^{-3} \tan({\mu\pi}/2) \big]}
\end{equation}
From (\ref{eq:15}), one has a non-phantom dark energy in the 
considered model, $-1 \leq \omega \leq 0$. This conclusion 
remains even if the independent barotropic fluids in the background
are taken into account. In that case, $\omega$ is replaced by the 
so-called effective EoS $\omega_{\text{eff}}$, defined as
\begin{equation}
  \label{eq:16}
  \omega_{\text{eff}} := \frac{p + \sum_{b}\omega_{b}\rho_{b}}{\rho +
    \sum_{b}\rho_{b}} =\frac{\omega 
    + \sum_{b} \omega_{b} \frac{\rho^{(0)}_{b}}{\rho} 
    (1+z)^{3(1+\omega_{b})}}{1+  \sum_{b}\frac{\rho^{(0)}_{b}}{\rho} 
    (1+z)^{3(1+\omega_{b})}}
\end{equation}
It follows from Eq.(\ref{eq:7}) that,
\begin{equation}
  \label{eq:17}
  \frac{\rho^{(0)}_{b}}{\rho} = \frac{2 \Omega_{b}^{(0)}}{\mu \pi \Omega^{(0)}
  } \arctan \Big[(1+z)^{-3}\tan({\mu\pi}/2) \Big]
\end{equation}
Hence,
\begin{equation}
  \label{eq:18}
  \omega_{\text{eff}} = \frac{\mu\pi \Omega^{(0)} \omega + 2 \arctan
    [(1+z)^{-3} \tan({\mu\pi}/2)] \sum_{b}\omega_{b}\Omega_{b}^{(0)}
    (1+z)^{3(1+\omega_{b})}}{\mu\pi \Omega^{(0)}  + 2 \arctan
    [(1+z)^{-3} \tan({\mu\pi}/2)] \sum_{b}\Omega_{b}^{(0)} 
    (1+z)^{3(1+\omega_{b})}}
\end{equation}
Cosmological acceleration occurs when $\omega_{\text{eff}} < -1/3$. By
assuming $\Omega^{(0)} \approx 0.73$, $\Omega_{m}^{(0)} \approx 0.27$, 
$\Omega_{r}^{(0)} \approx 8.1 \times 10^{-5}$ and $\mu\approx 0.931$, 
we plot Eqs.(\ref{eq:18}) and (\ref{eq:15}) in FIG.2, from which we see 
that $-1 \leq \omega_{\text{eff}} \leq 0$. In this model, $\omega_{\text{eff}}
< -1/3$ is possible only for $-2.6  \lesssim z \lesssim 0.6$. There are
two critical redshift points in the model under consideration, $z_{\text{c}}
\approx 0.6$ and $z_{\text{c}} \approx -2.6$. The former that corresponds
to the instant when the cosmological deceleration dominated by barotropic
matter ceases but the late time acceleration dominated by dark energy fluid 
starts is the past critical redshift. The latter is the future critical 
redshift, which predicts an incoming instant when the present cosmological acceleration 
dominated by dark energy fluid ceases but the decelerated expansion
dominated by barotropic matter recovers.
\begin{figure}[ht]
\begin{center}
\includegraphics{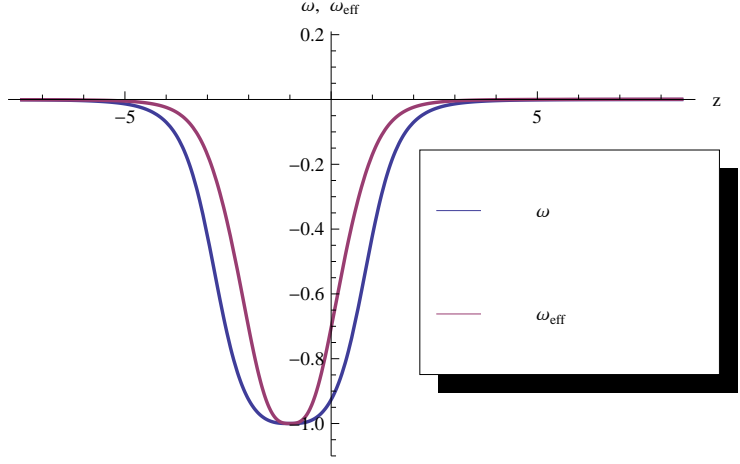}\hfill
\caption{\small{EoS of the suggested dark energy and effective EoS. We have
assumed that $\Omega^{(0)} \approx 0.73$, $\Omega_{m}^{(0)} \approx 0.27$,   
$\Omega_{r}^{(0)} \approx 8.1 \times 10^{-5}$ and $\mu\approx 0.931$. Both  
$\omega$ and $\omega_{\text{eff}}$ are bounded between $-1$ and $0$. 
$\omega_{\text{eff}} <- 1/3$ occurs when $-2.6 \lesssim z \lesssim 0.6$.}}
\label{hy2}
\end{center}
\end{figure}
When the universe enters into the acceleration phase can alternatively be
estimated by the deceleration parameter $q$, defined by
\begin{equation}
  \label{eq:19}
  q:= - \frac{\ddot{a}}{aH^{2}} = -1 + \frac{(1+z)}{H}\frac{dH}{dz}
\end{equation}
Substitution of Eqs.(\ref{eq:10}) and (\ref{eq:13}) into (\ref{eq:19}) gives,
\begin{eqnarray}
  \label{eq:20}
  &  & q= -1 + \frac{3 \arctan [(1+z)^{-3}\tan({\mu\pi}/2)]}{
    \mu\pi\Omega^{(0)} + 2 \arctan [(1+z)^{-3}\tan({\mu\pi}/2)] 
    \sum_{b} \Omega_{b}^{(0)} (1+z)^{3(1+\omega_{b})}} \nonumber \\
  &  & ~~~~~~~~~~~~~ \cdot \bigg \{ \frac{ \mu\pi \Omega^{(0)} 
  (1+z)^{3} \tan ({\mu\pi}/2)}{2 \arctan^{2}[(1+z)^{-3} \tan({\mu\pi}/2)]
  [(1+z)^{6} + \tan^{2}({\mu\pi}/2)]}  \nonumber \\
  &  & ~~~~~~~~~~~~~~~ + \sum_{b} (1+\omega_{b}) 
  \Omega_{b}^{(0)} (1+z)^{3(1+\omega_{b})}  \bigg \}\nonumber \\
  &  &
 \end{eqnarray}
The universe enters an acceleration phase when $q \leq 0$, which corresponds
to $-2.6 \lesssim z \lesssim 0.6$ when we assume $\Omega^{(0)} \approx 0.73$,
$\Omega_{m}^{(0)} \approx 0.27$, $\Omega_{r}^{(0)} \approx 8.1 \times 10^{-5}$
and $\mu\approx 0.931$.  See FIG.3 for detail. When $z$ approaches zero,
$q\approx -0.56$, which is in perfect agreement with the prediction ($q \approx
-0.55$) of $\Lambda\text{CDM}$ model of cosmology.  
\begin{figure}[ht]
\begin{center}
\includegraphics{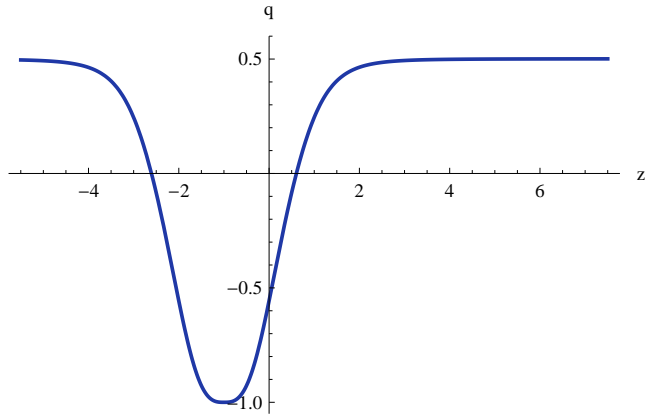}\hfill
\caption{\small{The deceleration parameter $q$ in the considered model, where
  we assume $\Omega^{(0)} \approx 0.73$, $\Omega_{m}^{(0)} \approx 0.27$,
  $\Omega_{r}^{(0)} \approx 8.1 \times 10^{-5}$ and $\mu\approx 0.931$. $q<0$
  occurs for $-2.6 \lesssim z \lesssim 0.6$.}}
\label{hy3}
\end{center}
\end{figure}

Let us now estimate the age of the universe in the present model. 
The cosmological time $t$ at any redshift $z$ is determined by 
formula
\[
t := \int\limits_{z}^{\infty} \frac{dz^{\prime}}{(1+z^{\prime}) H(z^{\prime})}
\]
The age $t_{0}$ of the universe is then obtained by setting $z=0$ 
in the above equation. For the model under consideration,
\begin{equation}
  \label{eq:21}
 t_{0} = \frac{1}{H_{0}}
  \int\limits_{0}^{\infty} \frac{d z^{\prime}}{ (1+z^{\prime}) \sqrt{
      \frac{\mu\pi \Omega^{(0)}}{2 \arctan[(1+z^{\prime})^{-3}\tan({\mu\pi}/2)]} 
      + \sum_{b} \Omega_{b}^{(0)}(1+z^{\prime})^{3(1+ \omega_{b})}}}
\end{equation}
We can integrate the integral in Eq.(\ref{eq:21}) numerically with the initial
conditions $\Omega|_{z=0}=\Omega^{(0)}$. The value of $t_0$ depends strongly 
upon what the fine-tuning parameter $\mu$ is assigned. We show this dependence 
in TABLE I by listing the possible values of $t_0$ for several values of 
parameter $\mu$, where we have taken $\Omega^{(0)} \approx 0.73$, $\Omega_{m}^{(0)}
\approx 0.27$, $\Omega_{r}^{(0)} \approx 8.1 \times 10^{-5}$ as before. To be 
consistent with the stellar age bound: $t_{0}> 11-12 ~\text{Gyr}$, the value 
of $\mu$ can not be too small ($\mu \gtrsim 0.738$). In this paper we choose $\mu 
\approx 0.931$, which leads to $t_{0}\approx 13~\text{Gyr}$. Hence the suggested 
dark energy model is free of the problem of universe age.
\begin{table}
\begin{center}
\begin{tabular}{|c|c|c|c|c|c|c|c|c|c|c|}
\hline
    $\mu$    & 0.70 & 0.73 & 0.738 & 0.75  & 0.80  & 0.85  & 0.90  & 0.931 & 0.95  & 0.98  \\
\hline
 $t_0$~(Gyr) & 10.76 & 10.95 & 11 & 11.08 & 11.48 & 11.95 & 12.55 &  13   & 13.31 & 13.85 \\
\hline
\end{tabular}
\end{center}
\caption{\small{The dependence of universe age upon the choice of the fine-tuning 
parameter $\mu$ in the suggested model, where we have taken $\Omega^{(0)} 
\approx 0.73$, $\Omega_{m}^{(0)} \approx 0.27$, $\Omega_{r}^{(0)} \approx 8.1 
\times 10^{-5}$. To satisfy the stellar bound on the age estimation, the magnitude
of $\mu$ has not to be less than 0.738.}}
\end{table}

\section{Dynamical mechanism}

The suggested fluid model of dark energy is consistent with the cosmological observation
data (at least at the background level). It provides a phenomenological explanation to 
the late time cosmic accelerated expansion and alleviates the coincidence problem to 
some extent. What is the dynamics behind such a phenomenological approach? From the
perspective of particle physics, the existence of a time dependent dark energy fluid 
implies that there is massive scalar field $\phi$ such as quintessence, phantom or 
tachyon coupled to gravity with a very flat scalar potential $V(\phi)$ \cite{Amendola1}. 
Because $-1\leq \omega \leq 0$, the suggested model might be associated with a tachyon 
field emerging from the scenario of superstring theory. The effective Lagrangian of 
a non-BPS D3-brane tachyon field $\phi$ reads,
\begin{equation}
  \label{eq:22}
  S = - \int d^{4}x V(\phi) \sqrt{-\det(g_{\mu\nu} + \partial_{\mu}\phi 
  \partial_{\nu}\phi)}
\end{equation}
where $V(\phi)$ is the tachyon potential. The energy momentum tensor determined
by the action (\ref{eq:22}) has the form
\begin{equation}
  \label{eq:23}
  T_{\mu\nu} = \frac{V(\phi) \partial_{\mu}\phi \partial_{\nu} \phi}{\sqrt{1+
      g^{\alpha \beta} \partial_{\alpha}\phi \partial_{\beta}\phi}} -
  g_{\mu\nu} V(\phi) \sqrt{1+ g^{\alpha\beta} \partial_{\alpha}\phi 
  \partial_{\beta}\phi}
\end{equation}
The energy density and the pressure density of this tachyon field are
$\rho_{\phi}=-T^{~0}_{0}$ and $p_{\phi}= \frac{1}{3}T^{~i}_{i}$. In a
flat FRW background it leads to
\begin{equation}
  \label{eq:24}
  \rho = \frac{V(\phi)}{\sqrt{1-\dot{\phi}^{2}}}, ~~~
     p = -V(\phi) \sqrt{1-\dot{\phi}^{2}}.
\end{equation}
It follows from Eq.(\ref{eq:24}) that EoS of the tachyon field has the form 
$p=-V^{2}(\phi)/\rho$. Hence the Chaplygin gas with EoS $p=-A/\rho$ (where 
$A$ is a positive constant) can be mimicked by a tachyon field with a constant 
potential. We now try to find the tachyon potential behind the suggested 
fluid model of dark energy with EoS given in Eq.(\ref{eq:4}). The Friedmann 
equation and the equation of state of the tachyon field are 
\begin{equation}
  \label{eq:25}
H^{2} = \frac{\rho}{3}= \frac{V(\phi)}{3\sqrt{1- \dot{\phi}^{2}}}, ~~~~
\omega_{\phi}= \frac{p}{\rho} = \dot{\phi}^{2} -1 .
\end{equation}
Manifestly, $-1 \leq \omega_{\phi} \leq 0$. By identifying $\omega_{\phi}$ and
$\rho_{\phi}$ with $\omega$ and $\rho$ given in Eqs.(\ref{eq:15}) and
(\ref{eq:7}) respectively, we get
\begin{equation}
  \label{eq:26}
  \frac{d\phi}{da} = \beta \sqrt{\frac{a}{1+ \alpha^{2} a^{6}}}
\end{equation}
where $\alpha = \tan(\mu \pi/2)$ and
\begin{equation}
  \label{eq:27}
  \beta = \frac{1}{H_{0}}\sqrt{\frac{2\alpha}{\mu\pi \Omega^{(0)}}}
\end{equation}
The solution of Eq.(\ref{eq:26}) which satisfies the initial condition
$\phi|_{a=0} \rightarrow 0$ reads,
\begin{equation}
  \label{eq:28}
  \phi(a) = \frac{2}{3} \beta a^{3/2} ~_{2}F_{1}\Big(\frac{1}{4}, \frac{1}{2};
  \frac{5}{4}; - \alpha^{2}a^{6} \Big)
\end{equation}
where $_{2}F_{1}(a, b; c; \zeta)$ is a hypergeometric function. The dependence
of scalar potential for the suggested dark energy tachyon field upon the scale
factor $a$ is found to be
\begin{equation}
  \label{eq:29}
  V(a) = \frac{3\mu\pi H_{0}^{2} \Omega^{(0)}}{2 \arctan (\alpha a^{3})}
  \sqrt{1- \frac{\alpha a^{3}}{(1+ \alpha^{2} a^{6}) \arctan (\alpha a^{3})}}
\end{equation}
Combination of Eqs.(\ref{eq:28}) and (\ref{eq:29}) enables us to further obtain
$V(\phi)$. It is difficult to have an analytical expression for $V(\phi)$ in 
the present case, we choose to plot the potential curve in FIG.4 instead.
\begin{figure}[ht]
\begin{center}
\includegraphics{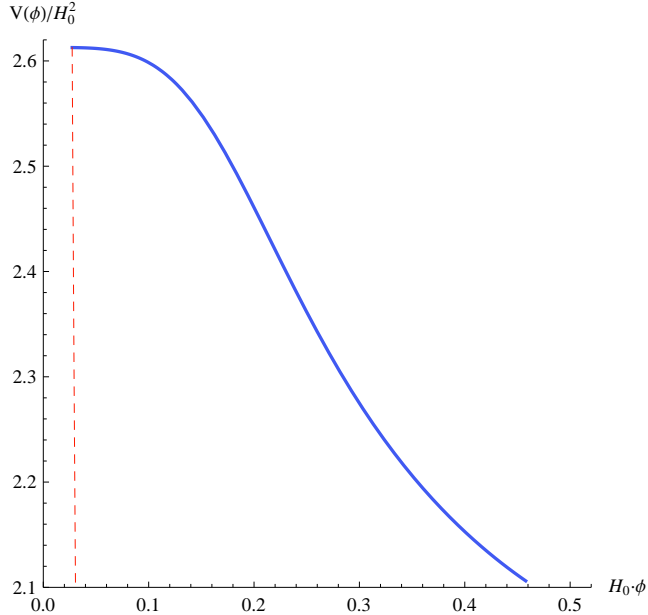}\hfill
\caption{\small The solid curve stands for the scalar potential of the tachyon field 
that mimics the suggested dark energy fluid, where $\Omega^{(0)} \approx 0.73$, 
$\mu\approx 0.931$, $\alpha \approx 9.19$ and $\beta \approx 0.093/H_{0}$. The dashed 
curve is a reference potential $V(\phi) \propto \phi^{-2}$ of power-law decelerated
expansion. The two potentials take the same initial value at $\phi \approx 0.03/H_0$. 
Obviously, the former is much flatter than the latter so that it can result in a late 
time accelerated cosmological expansion. }
\label{hy4}
\end{center}
\end{figure}
As expected, such a tachyon potential is not as steep as a reference potential $V(\phi) 
\propto \phi^{-2}$ of inverse square power-law of the power-law decelerated expansion
\cite{Aguirregabiria}. It is a competent tachyon potential responsible to an accelerated 
cosmological expansion.

\section{Discussion}
In this paper we propose a new fluid model of dark energy as an alternative to
the Chaplygin gas model. The energy density of dark energy fluid in the
suggested model is considerably suppressed in the barotropic matter dominant
eras, and it dominates the universe energy only for cosmic times of small
redshift. The model is in perfect agreement with the standard $\Lambda\text{CDM}$
model of cosmology on the present values of effective EoS of dark energy and 
deceleration parameter, and is free of the age problem of Big Bang model. To
understand the model in a dynamical manner, we interpret the fluid as a
tachyon field with a special potential, which is flatter than the inverse
square power-law potential of the decelerated power-law expansion. Phenomenologically, 
there are two differences between our model and $\Lambda\text{CDM}$ model in 
which the dark energy is viewed as a pure CC. The cosmic accelerated expansion will
continue forever in $\Lambda\text{CDM}$ model. In our model the universe is predicted to 
end up its present accelerated expansion in the near future (at $z\approx -2.6$) 
and reenter a decelerated expansion phase afterwards. Moreover, the severe 
coincidence problem of $\Lambda\text{CDM}$ model has been remarkably alleviated in 
our model. The investigation to this model in the present work is at the background 
level, it is unknown if there exists in the suggested model a strong integrated Sachs-Wolfe
(ISW) effect prohibiting the creation of CMB anisotropy. The study on curvature 
perturbations in the model is in progress, from which we will know whether the model
would be free of the pressure from CMB anisotropies.

\section*{Acknowledgement}
We would like to thank J. X. Lu, D. N. Gao, Y. J. Ouyang and M.L. Yan for valuable 
discussions. The work was supported in part by NSFC under No. 10375052.

\providecommand{\href}[2]{#2}\begingroup\raggedright\endgroup
\end{document}